# Mechanical properties of VMoNO as a function of oxygen concentration: toward development of hard and tough refractory oxynitrides

Running title: Mechanical properties of VMoNO as a function of oxygen concentration

Running Authors: Edström et al.

D. Edström, [1,2,a)] D. G. Sangiovanni, [1,3] L. Landälv, [1,4] P. Eklund, [1] J.E Greene, [1,5,6] I. Petrov, [1,5] L. Hultman, [1] and V. Chirita [1]

[1]Department of Physics, Chemistry, and Biology (IFM) Linköping University, SE 58183 Linköping, Sweden
[2]School of Science and Technology, Örebro University, SE-70182, Örebro, Sweden
[3]ICAMS, Ruhr-Universität Bochum, 44780 Bochum, Germany
[4]Sandvik Coromant AB, Stockholm SE-126 80, Sweden
[5]Frederick Seitz Materials Research Laboratory and the Materials Science Department,University of Illinois at Urbana-Champaign, Urbana, Illinois 61801, USA
[6]Materials Science Dept., National Taiwan Univ. Science & Technology, Taipei 10607, Taiwan.

[a)] Electronic mail: daniel.edstrom@liu.se

Improved toughness is a central goal in the development of wear-resistant refractory ceramic coatings. Extensive theoretical and experimental research has revealed that NaCl-structure VMoN alloys exhibit surprisingly high ductility combined with high hardness and toughness. However, during operation, protective coatings inevitably oxidize, a problem which may compromise material properties and performance. Here, we explore the role of oxidation in altering VMoN properties. Density functional theory and theoretical intrinsic hardness models are used to investigate the mechanical behavior of cubic $V_{0.5}Mo_{0.5}N_{1-x}O_x$ solid solutions as a function of the oxygen concentration x. Elastic-constant and intrinsic hardness calculations show that oxidation does not degrade the mechanical properties of $V_{0.5}Mo_{0.5}N$. Electronic structure analyses indicate that the presence of oxygen reduces the covalent bond character, which slightly lowers the alloy strength and intrinsic hardness. Nevertheless, the character of metallic d-d states, which are crucial for allowing plastic deformation and enhancing toughness, remains unaffected. Overall, our results suggest that VMoNO oxynitrides, with oxygen concentrations as high as 50%, possess high intrinsic hardness, while still being ductile.

## I. INTRODUCTION

During the past few decades, the development of thin films with increasingly higher hardness has been a key focus of materials science. [1] Hard coatings are typically deposited on industrial components such as tools, dies, engine parts, and architectural glass to protect parts from high operating temperatures, wear caused by extreme application conditions, oxidation, corrosion, and abrasion, thereby enhancing product lifetimes and consequently reducing operation costs. [2] Refractory transition-metal (TM) nitrides are commonly used as protective coatings on cutting tools and engine components due to their properties, which include high hardness [3–6] and wear resistance, [7] low coefficient of friction, [8] and high-temperature oxidation resistance. [9–11]

Increased hardness has been achieved using a variety of approaches, which primarily aim to reduce dislocation mobility by nanostructural engineering or tuning the concentration of point defects that act as pinning sites. Examples are synthesis of nano-scale composites, [12,13] superlattice structures, [14,15] and vacancy-induced hardening. [5,16,17] At a fundamental electronic level, thin-film hardness can be improved by tuning the valence electron concentration (VEC) of the material. Jhi et al. [18] demonstrated that a VEC of 8.4 corresponds to the maximum hardness in cubic transition-metal nitride and carbide alloys, as shear-resistant metal-N/C d-p electronic states are fully occupied, while metallic shear-sensitive d-d states remain empty. Hugosson et al. [19,20] employed VEC-tuning to set the cubic and hexagonal structures of transition-metal nitride and carbide alloys to equal formation energies. As the two structures are equally likely to be formed, the stacking fault concentration is maximized, which hinders slip across the faults and results in increased hardness.

However, a disadvantage of strategies based on increasing hardness by hindering dislocation motion is that they also decrease the ability of ceramics to dissipate stresses by plastic flow. [21] This often results in crack formation and propagation, which rapidly leads to brittle failure during use. Combining experiments, [6,22–24] *ab initio* calculations, [25–28] and density-functional molecular-dynamics simulations, [29] we have previously demonstrated by density functional theory (DFT) calculations that NaCl-structure (B1) $Ti_{0.5}Mo_{0.5}N$, $Ti_{0.5}W_{0.5}N$, $V_{0.5}Mo_{0.5}N$, and $V_{0.5}W_{0.5}N$ alloys are harder, as well as significantly more ductile (i.e. tougher), than their parent binary compounds TiN and VN. The enhanced toughness is induced by an optimized occupation of metallic d-d bonding states at the Fermi level, obtained upon the substitution of Mo/W atoms for Ti/V on the metallic (*Me*) sublattice. A higher VEC not only preserves the typically strong Me-N bonds found in TiN and VN, but also enables the formation of stronger Me-Me bonds compared to the parent compounds. Overall, high VEC lowers the shear resistance and promotes dislocation glide, thus leading to increased ductility. The effect was shown to be present regardless of ordering on the metallic sublattice. [25] In addition, the mixture of cubic- and hexagonal-structure binaries to form metastable B1 pseudobinary solid solutions (e.g., B1 TiN + hexagonal WN to form B1 $Ti_{0.5}W_{0.5}N$ [28]), was demonstrated to facilitate room-temperature $\{111\}\langle1\bar{1}0\rangle$ slip. Activating this slip system in addition to the more common $\{110\}\langle1\bar{1}0\rangle$ [30] system further enhances ductility and toughness.

Our previous studies were dedicated to understanding the mechanical behavior of refractory nitrides in oxygen-free conditions. However, thin films deposited in industrial systems often incorporate a substantial amount of oxygen [31] due to the relatively high base pressure at high-vacuum conditions, resulting in incorporation of oxygen primarily from residual water-vapor. Additionally, while TM nitrides are chemically stable at room temperatures, they oxidize readily at the elevated temperatures (> 1000 K) typically used in metal-cutting operations. [32] Moreover, oxynitrides are themselves interesting from a technological perspective. For example, it has been reported that TiAlNO/TiAlN multilayer structures exhibit higher wear resistance compared to conventional TiAlN coatings due to improved high-temperature properties resulting from increased oxidation resistance. [33] TiAlNO coatings have also been shown to be more corrosion-resistant in contact with molten aluminum alloys than TiN and TiAlN coatings [34], which is relevant for die casting. In addition, oxygen addition of up to 7 at.% in TiAlN coatings enhanced the wear resistance of TiAlN. [35] It is therefore of interest to investigate the effect of oxygen on the mechanical response and physical properties of TM nitrides.

Here, we use DFT *ab initio* modeling in an initial attempt at understanding the role of oxidation in altering the properties of stoichiometric $V_{0.5}Mo_{0.5}N$, which has been shown experimentally to be both hard and ductile. [22,24] We investigate the phase stability, elastic properties, intrinsic hardness, and electronic structure of B1 $V_{0.5}Mo_{0.5}N_{1-x}O_x$ solid solutions with x = 0.05, 0.09, and 0.50. This series of solid solutions serves as model reference system to probe the effects of oxygen substitution in the anion sublattice on film hardness and toughness. Our results indicate that $V_{0.5}Mo_{0.5}N$ mechanical properties are essentially unchanged by the inclusion of oxygen, thus providing a step toward development of tough and hard oxynitrides.

## II. MODELLING

We use the Vienna *ab initio* simulation package (VASP) [36] to carry out DFT calculations in the generalized gradient approximation of Perdew, Burke, and Ernzerhof (GGA-PBE). [37] Electron/ion interactions are described using projector augmented wave potentials (PAW). [38] Total energies are minimized to converge to within $10^{-5}$ eV/supercell using an energy cut-off of 500 eV for the plane-wave basis set. We calculate the mechanical properties of $VMoN_{1-x}O_x$ with $x = 0$, 0.05, 0.09, and 0.50, and the atoms arranged in a 128-atom special quasirandom structure (SQS). [39] The VMoN SQS-structure supercell is formed by interpenetrating Me and N sublattices with 64 fcc sites each, and non-orthogonal primitive vectors to minimize Me-Me correlations between neighboring shells; thus, mimicking disordered solid solutions. The oxygen-containing lattices are constructed starting from ideal SQS VMoN structures and replacing a sufficient number of randomly-selected N atoms with O atoms to reach the intended alloy composition. Structure optimizations are carried out by relaxing supercell volumes, shapes, and atomic positions using the conjugate gradient algorithm. [40] The Brillouin zone is sampled via the Monkhorst-Pack scheme [41] on 6x6x6 **k**-point grids.

Formation energies of B1 $V_{0.5}Mo_{0.5}N_{1-x}O_x$ alloys are calculated according to the expression

$$\Delta E_{V_{0.5}Mo_{0.5}N_{1-x}O_x} = E_{V_{0.5}Mo_{0.5}N_{1-x}O_x} - E_{comp}, \quad (1)$$

in which $E_{V_{0.5}Mo_{0.5}N_{1-x}O_x}$ is the energy per formula unit of the alloy, obtained directly from VASP output, and $E_{comp}$ is the minimum energy of the set of competing phases. $E_{comp}$ is determined by solving the linear programming equation,

$$\min E_{comp}(b^V, b^{Mo}, b^N, b^O) = \sum_i^n x_i E_i, \tag{2}$$

subject to the constraints

$$x_i \geq 0; \sum_i^n x_i^V = b^V, \quad \sum_i^n x_i^{Mo} = b^{Mo}, \quad \sum_i^n x_i^N = b^N, \quad \sum_i^n x_i^O = b^O. \tag{3}$$

Elastic constants $C_{11}$, $C_{12}$, and $C_{44}$, together with bulk moduli $B$, are obtained by fitting calculated energy versus strain $\delta$ curves, in which $|\delta| \leq 0.005$. Since the random alloys studied do not possess cubic symmetry, we use the projected cubic elastic constants, acquired following the method described in [25,42] as the averages

$$\bar{C}_{11} = \frac{C_{11} + C_{22} + C_{33}}{3}, \tag{4}$$

$$\bar{C}_{12} = \frac{C_{12} + C_{13} + C_{23}}{3}, \tag{5}$$

$$\bar{C}_{44} = \frac{C_{44} + C_{55} + C_{66}}{3}. \tag{6}$$

Isotropic elastic moduli, shear moduli, and Poisson ratios reported below are determined according to the Voigt-Reuss-Hill approach. [43]

We estimate intrinsic hardness using two methods. The first method is based on the empirical relationship suggested by Tian [44], which defines the intrinsic hardness as

$$H = 0.92 \left(\frac{G}{B}\right)^{1.137} G^{0.708}. \tag{7}$$

The shear modulus G and bulk modulus B are obtained from the elastic constants. The second method for estimating intrinsic hardness is an adaptation of the method proposed by Šimůnek. [45] Intrinsic hardness is expressed as

$$H_{ij} = \left(\frac{\beta}{\Omega}\right) S_{ij} e^{-\sigma f_{ij}}, \tag{8}$$

in which

$$f_{ij} = \left[\frac{e_i - e_j}{e_i + e_j}\right]^2. \tag{9}$$

Here, $\beta = 1550$ and $\sigma = 4$ are empirical constants, $\Omega$ the unit cell volume, $S_{ij}$ the bond strength between atoms $i$ and $j$, and $e_i = Z_i/R_i$ the reference energy of atom $i$. $Z_i$ is the valence of atom i and $R_i$ the radius at which the atom is electrically neutral; i.e. at which the integrated electron charge balances the charge of the nucleus. This expression is valid for perfect binary or single-element crystals and must be generalized for multicomponent materials in order to be applied to VMoNO.

A method for generalizing Eq. (8) to multicomponent materials was suggested by Hu et al. [46] in which the intrinsic hardness is expressed as

$$H_M = \left(\frac{\beta}{\Omega}\right) S_M e^{-\sigma f_M}, \tag{10}$$

for which $S_M$ and $f_M$ are the geometric averages of $S_{ij}$ and $f_{ij}$, respectively. The bond strength is calculated according to

$$S_{ij} = \sqrt{\left(\frac{e_i}{n_i}\right)\left(\frac{e_j}{n_j}\right)} \Big/ d_{ij}, \tag{11}$$

in which $d_{ij}$ is the bond length and $n_i$ the coordination number. In our application of this method, we use the average bond length obtained from VASP and a coordination number of six for all atoms. When calculating geometric averages, we first determine the number of first-neighbor bonds of each type, i.e V-N, V-O, Mo-N, and Mo-O, by analyzing the

VASP structure files.

# III. RESULTS AND DISCUSSION

## A. Formation energies

Table I lists the competing phases included in the formation-energy calculations and their respective energies.

TABLE I. Calculated energies per formula unit at T = 0 K for reference systems used in determining the formation energies of $V_{0.5}Mo_{0.5}N_{1-x}O_x$ as a function of x.

| Phase | Structure | Energy [eV/f.u.] | Phase | Structure | Energy [eV/f.u.] |
|-------|-----------|------------------|-------|-----------|------------------|
| V | BCC | -8.95 | $VO_2$ | Monoclinic | -25.7 |
| Mo | BCC | -10.9 | $V_2O_3$ | Trigonal | -43.8 |
| N | Free atom | -3.10 | $V_2O_5$ | Orthorhombic | -58.0 |
| O | Free atom | -1.60 | $V_3O_5$ | Monoclinic | -69.5 |
| $N_2$ | Dimer | -16.6 | $MoO_2$ | Tetragonal | -26.2 |
| $O_2$ | Dimer | -9.87 | $MoO_3$ | Orthorhombic | -32.9 |
| NO | Molecule | -12.3 | $V_{0.5}Mo_{0.5}N$ | Rocksalt SQS | -19.5 |
| VN | Rocksalt | -19.2 | $V_{0.5}Mo_{0.5}O$ | Rocksalt SQS | -17.6 |
| VO | Rocksalt | -16.7 | $VN_{0.5}O_{0.5}$ | Rocksalt SQS | -18.3 |
| MoN | Rocksalt | -19.2 | $MoN_{0.5}O_{0.5}$ | Rocksalt SQS | -18.7 |
| MoO | Rocksalt | -16.4 | | | |

Solving the optimization problem specified by Eqs. (2) and (3) yields the set of lowest-energy competing phases and their molar fractions, presented in Table II.

TABLE II. Solutions to the linear optimization problem in Equation 2 yielding the set of lowest-energy competing phases for each oxygen concentration x. Phases with concentrations found to equal zero are omitted from the table.

| | $x_{Mo}$ | $x_{V0.5Mo0.5N}$ | $x_{MoO2}$ | $x_{V2O3}$ |
|---|---|---|---|---|
| $V_{0.5}Mo_{0.5}N_{0.95}O_{0.05}$ | $1.88 \times 10^{-2}$ | $9.50 \times 10^{-1}$ | $6.20 \times 10^{-3}$ | $1.25 \times 10^{-2}$ |
| $V_{0.5}Mo_{0.5}N_{0.91}O_{0.09}$ | $3.38 \times 10^{-2}$ | $9.10 \times 10^{-1}$ | $1.12 \times 10^{-2}$ | $2.25 \times 10^{-2}$ |
| $V_{0.5}Mo_{0.5}N_{0.5}O_{0.5}$ | $1.88 \times 10^{-1}$ | $5.00 \times 10^{-1}$ | $6.25 \times 10^{-2}$ | $1.25 \times 10^{-1}$ |

For all three oxygen concentrations, the most important competing phases are BCC Mo, NaCl-structure $V_{0.5}Mo_{0.5}N$, tetragonal P42/mnm $MoO_2$, and trigonal $R\bar{3}c$ $V_2O_3$. Calculated formation energies of $V_{0.5}Mo_{0.5}N_{1-x}O_x$ are presented as a function of x in Figure 1.

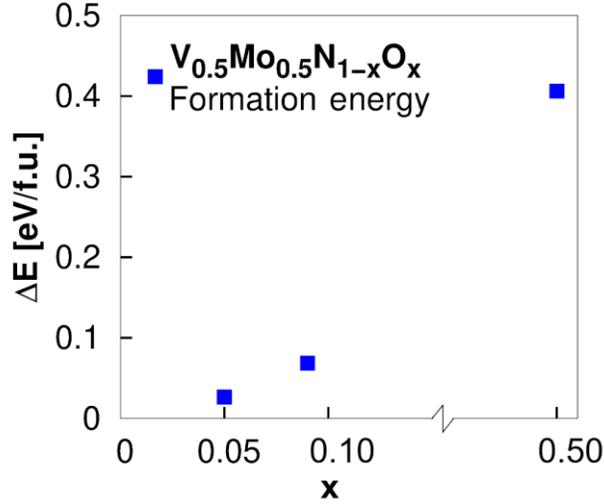

FIG. 1. Formation energies of $V_{0.5}Mo_{0.5}N_{1-x}O_x$ versus the oxygen fraction, x = O / (N + O), on the anion sublattice.

The results show that the formation energy is positive for all oxygen fractions investigated, which indicates that the $V_{0.5}Mo_{0.5}N_{1-x}O_x$ alloy is thermodynamically unfavored at T = 0 K. At finite temperatures, however, the configurational entropy – and likely also the vibrational entropy [47] – terms will result in reductions in the alloy free energy of mixing. Since the formation energies of $V_{0.5}Mo_{0.5}N_{0.95}O_{0.05}$ and $V_{0.5}Mo_{0.5}N_{0.91}O_{0.09}$ are less than 100 meV/f.u., entropy contributions at elevated temperatures are expected to stabilize alloys with relatively low oxygen contents. In contrast, the formation energy of B1 $V_{0.5}Mo_{0.5}N_{0.5}O_{0.5}$ is strongly positive (~400 meV/f.u.). However, a high oxygen content does not exclude the possibility of a kinetically-stabilized B1 structure, as shown by the arc-evaporation deposition of TiAlON thin films containing up to 35 at.% oxygen [35]. Furthermore, we note that the calculations performed here represent a simplified model for evaluating trends with respect to oxygen content. In reality, a high oxygen concentration may induce phase transitions and/or introduce defects. Taking these possibilities into account is beyond the scope of the present work.

## B.  Mechanical properties

Calculated lattice constants and elastic properties of $V_{0.5}Mo_{0.5}N_{1-x}O_x$ alloys are presented in Table III.

TABLE III. Calculated $V_{0.5}Mo_{0.5}N_{1-x}O_x$ alloy lattice constants, $a$, and mechanical properties. $B$, $E$, and $G$ are the bulk, elastic, and shear moduli; $v$ is Poisson's ratio; and $\bar{C}_{44}$, $\bar{C}_{11}$, and $\bar{C}_{12}$ are the fundamental cubic elastic constants.

|  | $a$ [Å] | $B$ [GPa] | $E$ [GPa] | $G$ [GPa] | $v$ |
|---|---|---|---|---|---|
| $V_{0.5}Mo_{0.5}N$ | 4.241 | 303 | 293 | 109 | 0.393 |
| $V_{0.5}Mo_{0.5}N_{0.95}O_{0.05}$ | 4.245 | 296 | 279 | 104 | 0.395 |
| $V_{0.5}Mo_{0.5}N_{0.91}O_{0.09}$ | 4.247 | 292 | 272 | 101 | 0.396 |
| $V_{0.5}Mo_{0.5}N_{0.5}O_{0.5}$ | 4.325 | 217 | 207 | 77 | 0.394 |
|  | $\bar{C}_{44}$ [GPa] | $\bar{C}_{11}$ [GPa] | $\bar{C}_{12}$ [GPa] | $\bar{C}_{12} - \bar{C}_{44}$ [GPa] | |
| $V_{0.5}Mo_{0.5}N$ | 87 | 508 | 200 | 114 | |
| $V_{0.5}Mo_{0.5}N_{0.95}O_{0.05}$ | 85 | 482 | 203 | 118 | |
| $V_{0.5}Mo_{0.5}N_{0.91}O_{0.09}$ | 83 | 473 | 201 | 118 | |
| $V_{0.5}Mo_{0.5}N_{0.5}O_{0.5}$ | 76 | 322 | 164 | 89 | |

The crystal structure expands upon introducing oxygen into the lattice, corresponding to an increase in the lattice parameter of ~2% with x = 0.5. Introduction of oxygen on the anion sublattice is also likely to introduce vacancies on the metal sublattice due to charge balancing, [48] which may have significant effects on material properties. The initial results presented here should thus be interpreted as trends for the strengths of metal-oxygen and metal-nitrogen bonds.

We find that $V_{0.5}Mo_{0.5}N_{1-x}O_x$ elastic properties are largely unchanged for small concentrations of oxygen (x = 0.05, 0.09). The primary differences are in the elastic moduli E and the $C_{11}$ elastic constants, which are reduced from 293 and 508 GPa at x = 0 to 272 and 473 GPa at x = 0.09, respectively. However, with x = 0.50, B, E, and G decrease by ~30%. The elastic constants also decrease with the largest reduction, ~37%, occurring in $C_{11}$.

The ductility of materials systems can be qualitatively evaluated using the empirical criteria of Pugh [49] and Pettifor [50], which associate ductile behavior with G/B < 0.5 and a Cauchy pressure $(C_{12} - C_{44}) > 0$, respectively. Niu et al. [51], who generalized the Pettifor and Pugh criteria to different classes of materials, showed that an increasing G/B ratio is indicative of enhanced strength, while an increasing $(C_{12}–C_{44})$/E ratio is associated with increased ductility. According to the latter criteria, visualized in Figure 2, the strength of VMoNO solid solutions decreases slightly with increasing O content, while the ductility remains unchanged.

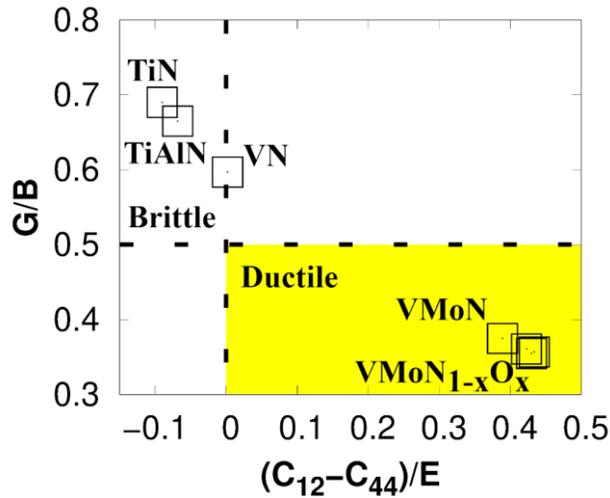

FIG. 2. Calculated shear-to-bulk moduli (G/B) ratios versus the ratio of the Cauchy pressure to the elastic modulus, $(C_{12}-C_{44})$/E, for $V_{0.5}Mo_{0.5}N_{1-x}O_x$ with x = 0, 0.05, 0.10, and 0.50. Values are also shown for the reference compounds TiN and VN and alloy $Ti_{0.5}Al_{0.5}N$.

The Poisson ratio ν, also used as an indication of ductility, [52] supports the validity of Niu's criteria as it remains essentially constant as x is varied from 0 to 0.5. The reductions in B, G, E, and $C_{11}$ observed with x = 0.50 (Table III) also suggest a decrease in alloy strength with significant incorporation of oxygen. For comparison, we include TiN, VN, [22] and $Ti_{0.5}Al_{0.5}N$ [26], all known to be brittle, in Figure 2.

The intrinsic hardnesses of $V_{0.5}Mo_{0.5}N_{1-x}O_x$ alloys, estimated using the method of Tian et al (Eq. 7) , $H_{Tian}$, as well as that of Hu et al (Eq. 10), $H_{Hu}$, are presented in Table IV.

TABLE IV. Calculated intrinsic hardness H based upon the model of Tian et. al, $H_{Tian}$, and the generalized Šimůnek model of Hu et. al, $H_{Hu}$.

| Alloys | $H_{Tian}$ [GPa] | $H_{Hu}$ [GPa] |
|---|---|---|
| $V_{0.5}Mo_{0.5}N$ | 7.37 | 22.3 |
| $V_{0.5}Mo_{0.5}N_{0.95}O_{0.05}$ | 7.51 | 21.7 |
| $V_{0.5}Mo_{0.5}N_{0.91}O_{0.09}$ | 7.22 | 21.7 |
| $V_{0.5}Mo_{0.5}N_{0.5}O_{0.5}$ | 6.13 | 18.6 |

The two methods produce quite different results. For $V_{0.5}Mo_{0.5}N$, $H_{Tian}$ = 7.37 while $H_{Hu}$ = 22.3. We note that experimentally, the hardness of $V_{0.5}Mo_{0.5}N$ has been measured as 20 GPa [22], which is in reasonable agreement with $H_{Hu}$. Calculated $V_{0.5}Mo_{0.5}N_{1-x}O_x$ intrinsic hardnesses vs. x, obtained using Hu's method, are plotted in Figure 3 together with the experimental hardness of $V_{0.5}Mo_{0.5}N$ [22].

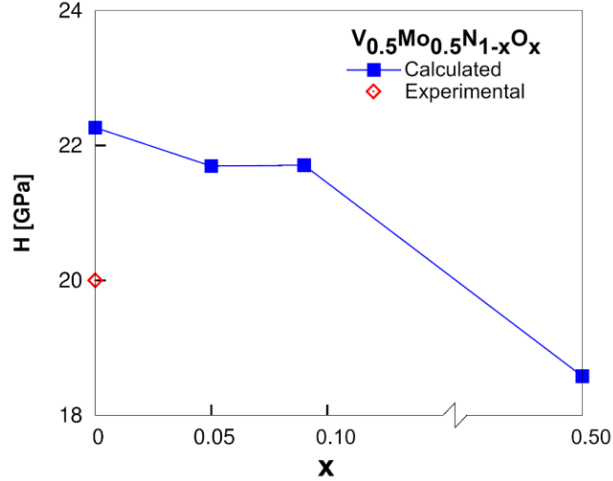

FIG. 3. Calculated intrinsic hardnesses H, using the extended Simunek method of Hu, [49] for $V_{0.5}Mo_{0.5}N_{1-x}O_x$ as a function of x = O / (N + O). The experimentally determined hardness of $V_{0.5}Mo_{0.5}N$ is also shown at x = 0. [22]

While Tian's approach provides intrinsic hardness values which are far too low, the trends as a function of O concentration are the same as those of Hu, which shows that H decreases only slightly when up to 10% of the nitrogen is replaced by oxygen. At x = 0.50, the intrinsic hardness decreases by ~17%. Based on our calculated formation energies it is unlikely to reach such high oxygen concentrations, and consequently the film properties are unlikely to degrade significantly due to oxygen incorporation.

## C.  *Charge densities*

Figure 4 shows calculated charge densities for fully-relaxed $V_{0.5}Mo_{0.5}N_{0.95}O_{0.05}$ and $V_{0.5}Mo_{0.5}N_{0.5}O_{0.5}$, together with charge densities after undergoing 10% tetragonal and trigonal strain.

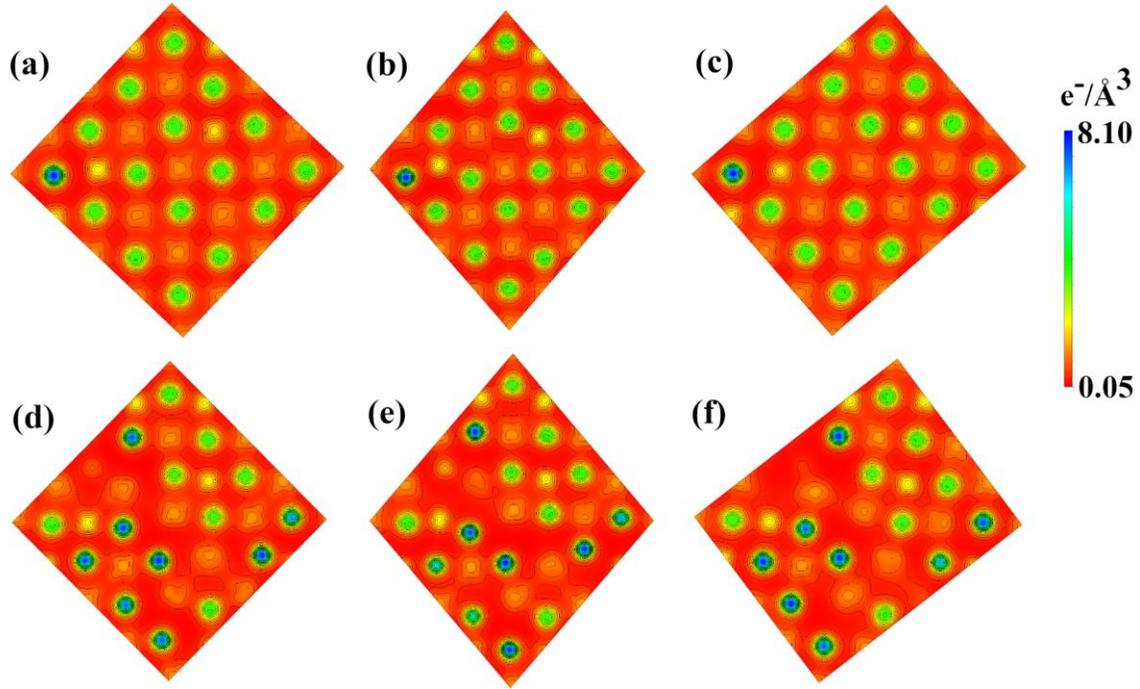

FIG. 4. Charge-density maps for $V_{0.5}Mo_{0.5}N_{0.95}O_{0.05}$ (upper row) and $V_{0.5}Mo_{0.5}N_{0.5}O_{0.5}$ (lower row) with (a), (d) 0% strain; (b), (e) 10% tetragonal strain; and (c), (f) 10% trigonal strain. The charge density color code is shown on the right side of the figure.

Oxygen atoms, easily identified by their high charge densities, appear blue in the images, whereas nitrogen atoms appear as green circles (see color scales vs. electron density). The charge density in metal/non-metal bonds is reduced near oxygen atoms (i.e., the bonds are weakened) in both the relaxed and strained structures. As expected, it is also clear that the distortion of the relaxed crystal structure is much more pronounced with 50% oxygen on the anion sublattice than with 5%. Overall, it is primarily the covalent bonds which are affected. The N and metal charge concentrations do not change significantly with the addition of oxygen. However, as previously noted, this amount of oxygen would likely induce the creation of metal vacancies due to charge balancing which could alter the charge densities.

# IV. SUMMARY AND CONCLUSIONS

We have calculated formation energies and mechanical properties of $V_{0.5}Mo_{0.5}N_{1-x}O_x$ alloys as a function of x = 0, 0.05, 0.09, and 0.50. The formation energies obtained indicate that the oxynitride alloys are thermodynamically unfavorable at T = 0 K. However, configurational and vibrational entropy terms at finite temperatures may stabilize alloys with x = 0.05 and 0.09, for which the substitution of oxygen on the nitrogen sublattice has limited effects on the mechanical properties. However, at x = 0.50, bulk, shear, and elastic moduli $B$, $G$, and $E$ are reduced by 30%. $V_{0.5}Mo_{0.5}N$ has been shown experimentally to be ductile [22]. Based upon the ductility criteria of Pugh and Pettifor, the $V_{0.5}Mo_{0.5}N_{1-x}O_x$ alloys remain ductile with x up to 0.5. We have estimated the intrinsic hardness of $V_{0.5}Mo_{0.5}N_{1-x}O_x$ using two different methods, which provide trends as a function of x that agree qualitatively, and yield a intrinsic hardness reduction of only 17% at x = 0.50. Of the two intrinsic hardness calculation approaches, the method proposed by Hu [46] matches $V_{0.5}Mo_{0.5}N$ experimental results much better than that of Tian et al; [44] $H_{Hu}$ = 22.3 GPa at x = 0 and 18.6 GPa at x = 0.50.

## ACKNOWLEDGMENTS

We acknowledge financial support from the Swedish Government Strategic Research Area Grant in Materials Science on Advanced Functional Materials at Linköping University (Faculty Grant SFO-Mat-LiU No. 2009 00971). Calculations were performed using the resources provided by the Swedish National Infrastructure for Computing (SNIC), on the Triolith and Tetralith Clusters located at the National Supercomputer Centre (NSC) in Linköping, and on the Kebnekaise cluster located at the High Performance Computing Center North (HPC2N) in Umeå. D.G.S. gratefully acknowledges financial support from the Olle Engkvist Foundation and the competence center FunMat-II supported by the Swedish Agency for Innovation Systems (Vinnova, grant no 2016-05156).

## Figure Captions

FIG. 1. Formation energies of $V_{0.5}Mo_{0.5}N_{1-x}O_x$ versus the oxygen fraction, $x = O / (N + O)$, on the anion sublattice.

FIG. 2. Calculated shear-to-bulk moduli (G/B) ratios versus the ratio of the Cauchy pressure to the elastic modulus, $(C_{12}-C_{44})/E$, for $V_{0.5}Mo_{0.5}N_{1-x}O_x$ with $x = 0$, 0.05, 0.10, and 0.50. Values are also shown for the reference compounds TiN and VN and alloy $Ti_{0.5}Al_{0.5}N$.

FIG. 3. Calculated intrinsic hardnesses H, using the extended Simunek method of Hu, [49] for $V_{0.5}Mo_{0.5}N_{1-x}O_x$ as a function of $x = O / (N + O)$. The experimentally determined hardness of $_{V0.5}Mo_{0.5}N$ is also shown at $x = 0$. [22]

FIG. 4. Charge-density maps for $V_{0.5}Mo_{0.5}N_{0.95}O_{0.05}$ (upper row) and $V_{0.5}Mo_{0.5}N_{0.5}O_{0.5}$ (lower row) with (a), (d) 0% strain; (b), (e) 10% tetragonal strain; and (c), (f) 10% trigonal strain. The charge density color code is shown on the right side of the figure.